# Chapter 3

## Credibility Analysis in Social Big Data

Bilal Abu-Salih[1], Pornpit Wongthongtham[2]
Dengya Zhu[3] , Kit Yan Chan[3] , Amit Rudra[3]

[1]The University of Jordan
[2]The University of Western Australia
[3]Curtin University

**Abstract:** The concept of social trust has attracted an attention of information processors/data scientists and information consumers / business firms. One of the main reasons for acquiring the value of SBD is to provide frameworks and methodologies using which the credibility of online social services users can be evaluated. These approaches should be scalable to accommodate large-scale social data. Hence, there is a need for well comprehending of social trust to improve and expand the analysis process and inferring credibility of social big data. Given the exposed environment's settings and fewer limitations related to online social services, the medium allows legitimate and genuine users as well as spammers and other low trustworthy users to publish and spread their content. This chapter presents an overview of the notion of credibility in the context of SBD. It also list an array of approaches to measure and evaluate the trustworthiness of users and their contents. Finally, a case study is presented that incorporates semantic analysis and machine learning modules to measure and predict users' trustworthiness in numerous domains in different time periods. The evaluation of the conducted experiment validates the applicability of the incorporated machine learning techniques to predict highly trustworthy domain-based users.

## 3.1 Introduction

The dissemination of information via the World Wide Web is no longer a monopoly; anyone can be a producer and a publisher of information. Given the abundance of data, it is hard for those who receive such quantities of information to distinguish the accurate from the inaccurate, the good from the poor. With the development of online social services, and the huge amount of social data that is generated, the quality of social data has not improved; rather, all types of false information have permeated these platforms. Rumours, for instance, are an example of bad quality data [1]. Rumours are a negative social phenomenon that is prevalent in societies. It is also one of the most serious psychological and moral wars that are raging in an atmosphere charged with various economic, political and social factors [2]. Spam is another category of low-quality content. Social spam content such as fake accounts, bulk messaging (*sending the same post many times in a relatively short period of time*), malicious links, fake reviews, etc. have degraded the quality of experience obtained by the social community members [3].

Data credibility varies according to the reputation of the data producer. For example, in online social services, all users' posts do not have the same level of reputation; a tweet from a verified user who has established a broad audience of followers has more impact than a tweet from a new user or a user with a small number of followers. Producers of bad quality social data provide their content via text, sound, image, and video which allow them to proliferate, especially since they can do so with anonymity and impunity. Due to the huge amount of information flowing to its recipients, in conjunction with the lack of a gatekeeper for those sites, it is difficult to verify the content, thereby making it easier for others to perform the task of disinformation [4]. Thus, online social services are hijacked, and their otherwise useful tools are misused to create chaos and spread false news, and to undermine intellectual convictions, ideological constants, and moral and social factors that could cause confusion within the community. This has become a threat to social security and social harmony as a result of the absolute freedom guaranteed by those sites [5-8].

On the other hand, the good quality content obtained from SBD has several significant impacts [9]. The use of social media is an empowering force in the hands of the public and private sectors and can have a positive influence on a community's development. It is an important tool to spread (public health) awareness, ensure security, and improve social and economic practices. Online social services consolidate and strengthen relationships between the users by sharing factual information and exchanging views on a variety of topics. This gives individuals considerable experience in many domains, in addition to enabling them to acquire knowledge and skills. Furthermore, the extraction and examination of quality content can benefit several vital sectors. For example, high-quality social data leads to a better understanding of customer behaviour and keeps a company's audience updated with the latest developments which improve customers' experience and increases revenues [10]. Last but not least, the quality of data influences the decision-making process of business operators [11, 12].

Data credibility varies according to the reputation of the data producer. For example, in online social services, all users' posts do not have the same level of reputation; a tweet from a verified user who has established a broad audience of followers has more impact than a tweet from a new user or a user with a small number of followers. Producers of bad quality social data provide their content via text, sound, image, and video which allow them to proliferate, especially since they can do so with anonymity and impunity. Due to the huge amount of information flowing to its recipients, in conjunction with the lack of a gatekeeper for those sites, it is difficult to verify the content, thereby making it easier for others to perform the task of disinformation [4]. Thus, online social services are hijacked, and their otherwise useful tools are misudused to create chaos and spread false news, and to undermine intellectual convictions, ideological constants, and moral and social factors that could cause confusion within the community. This has become a threat to social security and social harmony as a result of the absolute freedom guaranteed by those sites [5-8].

On the other hand, the good quality content obtained from SBD has several significant impacts [9]. The use of social media is an empowering force in the hands of the public and private sectors and can have a positive influence on a community's development. It is an important tool for creating a better future by harnessing these platforms to spread (public health) awareness, ensure security, and improve social and economic practices. Online social services consolidate and strengthen relationships between the users by sharing factual information and exchanging views on a variety of topics. This gives individuals considerable experience in many domains, in addition to enabling them to acquire knowledge and skills. Furthermore, the extraction and examination of quality content can benefit several vital sectors. For example, high-quality social data leads to a better understanding of customer behaviour and keeps a company's audience updated with the latest developments which improve customers' experience and increases revenues [10]. Last but not least, the quality of data influences the decision-making process of business operators [11, 12].

SBD analytics provides advanced technical capabilities to the process of analysing massive and extensive social data to achieve cavernous insights in an efficient and scalable manner. With such increase in volume and diversity of data that businesses are dealing with today, they found themselves at a crossroad; either ignoring these data, or gradually starting to adapt, understand and benefit from them[13, 14]. Hence, to efficiently incorporate data analytics to benefit the organisations, customers data should be collected directly (i.e. internal operations collected from day-to-day transactional information systems) and indirectly(i.e. social data collected from online social services platforms). The data collected from online social services should be examined to infer high-quality content and eliminate the poor quality data for further data analysis. To obtain this objective, organisations should understand how to handle the propagation and the heterogeneous of the BD. Thus they should address the following two main concerns: (i) BD are huge in volume, varies in nature, and sprawl rapidly; (ii) Extracting valuable and veracious data is a key challenge. Hence, there is a need for an approach for data analytics capable of handling the BD features and proficiently able to filter out unsolicited data and infer a value. This approach should comprise an advanced technical solution able to capture such huge generated data, scrutinising the collected data to eliminate unwanted data, measure the quality of the inferred data, and transform the amended data to benefit further data analytics techniques.

This chapter provides an overall depiction to credibility in the context of SBD. It also lists some of the recently developed approaches to address this problem. In particular, the next section discusses the credibility problem in the context of Social Big Data. Section 3 list some of the approaches followed to address credibility in its generic essence as well as tackling domain specific credibility. Section 4 provides an overview of an advanced approach designed to address the domain-based credibility in SBD. A detailed description is given of the overall mechanism of the developed approach and includes a metric of the key attributes used to measure the trustworthiness of domain-based users. Finally, this chapter concludes with a summary to main aspects discussed in this chapters and suggestions that should be made in the future.

## 3.2 An Overview of Credibility in SBD

Social Trust or Social Credibility, in information science, is the measurement of confidence where a group of individuals or communities behave predictably. It can be described through offering reasonable grounds for being believed [15, 16]. Adding a user-domain dimension when calculating trust in social media is an important factor. This helps to enhance understanding of users' interest. In this context, the notion of domain-based trust for the data extracted from the unstructured content (such as social media data) is significant. This is through calculating trustworthiness values which correspond to a particular user in a particular domain.

**Problem Statement:** *Lack of domain-based credibility approaches in the era of SBD, refers to the deficiency of implementing platforms to measure and evaluate the credibility of users in SBD considering their domain(s) of interest. These platforms should address the BD features, and facilitate data storage, data processing and data analysis.*

As discussed in Chapter 2, and despite the diverse depictions of the Big data problem, Big data is known to be described through several **V**-features. These include, and are not limited to: *Volume* refers to the vast increase in the data growth; *Velocity* represents the accumulation of data in high speed and real-time from several data sources; *Variety* involves fuzzy and heterogeneous types of data; *V*eracity refers to the accuracy, correctness and trustworthiness of data; *Variability* refers to variance in meaning[17]; and *Value* represents the outcome product of Big data analysis(i.e. new insights)[18]. Within this depiction, the impact of Big data abundance extends beyond business-related data to cope political and governmental data, healthcare data, education-related data, and many other industries. The key challenge of Big data analysis is the mining of enormous amounts of data in the quest for added value.

Online social services is a primal Big data island provides a momentum dense of social data which require a deep scrutinising. These web social networks have thrown open the doors of platforms for people to unleash their opinions and build new varieties of social communications based on these virtual societies. Online social services provide fertile mediums for legitimate users as well as spammers to publish their content. The spamming activities in social platforms increased dramatically [19]. The spammer's activities comprise abuse in utilising online social services' features and tools; spammers send annoying messages to legitimate users; their contents include malicious links, and hijack popular topics [20]. Spammers post contents for various topics, and they duplicate posts [19]. Further, to propagate their vicious activities, spammers abuse other online social services' features such as hashtags, and mention other users and Link-shortening services [21]. Hence, it is important to understand the users' behaviour because of the dramatic increase in the usage of online social platforms. For example, since there are over 330 million monthly active users of Twitter[1] , a significant question arises regarding the quality of the enormous data that is being proliferated every minute by users of such computer-generated environments. This explains the importance of measuring the users' credibility and ascertaining the users' influence in a particular domain. Hence, the factual grasp of the users' domains of interest and an appropriate judgement of their emotions enhances the customer-to-business engagement. This necessitates an accurate analysis of customer reviews and their opinions to obtain a better understanding of their needs thereby enhancing their customer service.

Despite the considerable efforts conducted to untangle the social trust problem; However, there are still vital dimensions need to be addressed to consolidate the proposed approaches. This includes tracking and monitoring users trustworthiness overtime; improve the existing semantic analysis techniques toward further enhancement to the contextual understanding to the users textual content; further incorporation to the sentiment analysis methods to listen to the voice of the user's followers and their opinions toward him/her. Last but not least, the implementation of the proposed approaches should address the key feature of Big data and provides technical solutions to the massive data generated steadily and incessantly from the online social platforms. The next section provides an overview of solutions that are proposed to measure users credibility in online social services.

## 3.3 Credibility Approaches in SBD

In modern enterprises, social networks are used as part of the infrastructure for a number of emerging applications such as recommendation and reputation systems [22-35]. In such applications, trust is one of the most important factors in decision making. The significance of trust is evident in multiple disciplines such as computer science, sociology, and psychology. Trust evaluation in the social media ecosystem is still immature; hence extensive research is required in this area [15]. The current literature has shown an ongoing enhancement to the proposed approaches to measure, evaluate and quantify the trustworthiness values inferred from the users of OSNs and their content. These approaches can be divided into two main categories conducted within the context of SBD; (i) proposed solutions to address the problem of generic social trust; and (ii) proposed solutions to infer domain-driven social trust.

---

[1] https://www.wordstream.com/blog/ws/2020/04/14/twitter-statistics , Accessed 21 04. 2020

### 3.3.1 Generic-based Trustworthiness Approaches

Generic-based trust approaches in OSNs are those frameworks, techniques, and tools developed to calculate and infer the trustworthiness values of users and/or their content with no consideration to the domain(s) of interest which can be extracted from the user level or post level. The trustworthiness of social media data is now a crucial consideration. With such a vast volume of data being interchanged within the social media ecosystems, data credibility is a vital issue, especially regarding personal data [36]. There are some approaches to measuring trustworthiness in social media [37-45] [46] [47]. Nepal et al. [48] address the challenges of the trust in the web-based social media. The paper states the four main components that are involved in the trust evaluation; service consumers, service providers, services and content. As indicated in their paper, the key challenges in social trust context are: (i) how to assign a trust value for a new entity (user/post)?(ii) trust propagation issue in OSNs;(iii) trust for recommender systems;(iv) addressing the trickery phenomena in the social web environment(wrong content, personal identity and location).

**Graph-based Social Trust:** Podobnik et al. [43] propose a model that calculates trust between friends in a network graph based on weights of the edges between user's connected friends in Facebook. Agarwal and Bin [42] suggest a methodology to measure the trustworthiness of a social media user by using a heterogeneous graph in which each actor in the Twitter domain was presented as a vertex type in the graph. The level of trustworthiness was measured using a regressive spread process. The paper, on the other hand, neglects to consider the importance of weighting scheme and time factor. Each edge category should be assessed at different credibility levels; hence, a weighting scheme should be used. Trustworthiness values differ over time; consequently, the temporal/time factor should be integrated. Authors of [49] propose trust propagation scheme to predict consumer's trust value on the service provider in service-oriented social networks considering structural properties of social networks and exploits the association between degree distribution and trust distribution for performance optimization purpose. TweetCred [50] is a Support Vector Machine credibility ranking inference framework for Twitter users in real-time data motion. Abbasie et al.[51] present an algorithm called CredRank to cluster users of the social media based on their online behaviour to detect the coordinated users. Authors of [52] describe a new methodology for post reputation based on weighted social interaction (in Twitter). Jeong et al. [47] discuss the perspectives of followers to a specific followee in the Twitter domain; The paper classifies the followers in three clusters based on their feelings(support, non-support and neutrality). Naumann [53] discusses the relationship between the message aim or intention of the user and his level of trust in the company. The paper answered the question of the effect of the message intention on the trust variable in the domain of B2B companies. Kopton et al. [54] explore away to evaluate trust via functional magnetic resonance imaging (fMRI) for better understanding users' behaviour on the OSNs; the idea is to study brain activity for users when they are engaged with the social platforms. Authors of [55] incorporate a number of attributes; indegree(#followers), retweets, and mentions to measure users' trustworthiness.

Brown et al. [56] adopt K-shell algorithm to measure users influence. The algorithm takes a graph of followers/following relationship as input and evaluates the k-shell level which forms users' ranking. Arlei et al. [38] examine the influence of social media users and the significance of their contents in information dissemination data. Work of Tsolmon and Lee's [41] means to measure the trustworthiness of Twitter users. Parameters of the Following-Ratio (#follower/#following) and Retweet-Ratio (total number of retweets of user/total number of tweets) are harnessed to infer well-known users using the HITS Algorithm mechanism. Cutillo et al. [57] present a technique to confirm the privacy of the OSNs' users using a new method to handle certain security and privacy issues.

**Trust for recommendation systems:** In [58], the authors show a web of trust as an alternative to the standard way of ranking a user, i.e. standard recommendation systems. Further, Gupta et al. [59] present the "WTF: Who To Follow" service which is being used as a recommendation system for the Twitter social network. This service is used mainly as a recommendation driver and has a significant impact; by using it, numerous new connections have been created. Further work by [60] and [61] propose trusted-based recommendation techniques.

**Trust incorporating Sentiment Analysis**: The use of sentiment analysis techniques to analyse the content of online social services has significantly influenced several aspects of research. In the context of social trust, authors of [62] propose a recommendation system framework incorporating implicit trust between users and their emotions. AlRubaian et al. [63] present a multi-stage credibility framework for the assessment of microbloggers' content. The development of sentiment-based trustworthiness approaches for online social services is discussed further in [64-66].

**Business Intelligence Incorporating Trust:** Business Intelligence applications are more focused on structured data; however, to understand and analyse the social trust, there is a need to collect data from various sources. Collective intelligence has spread to many different areas, with a particular focus on fields related to everyday life such as commerce, tourism, education, and health, causing the size of the social Web to expand exponentially. SBD exhibits all the typical properties of big data: wide physical distribution, diversity of formats, non-standard data models, independently-managed and heterogeneous semantics. Labrinidis and Jagadish [67] summarised the following challenges of Big data 1. Data acquisition (infers useful data and discard irrelevant). 2. Building the right metadata for data description. 3. Data extraction and formation. 4. Data quality (value). 5. Automatic data analysis. 6. Coordination between traditional SQL with NoSQL methods.

The incorporation of Big data technology to facilitate data analysis tools is considered to be a hot topic, especially regarding the contents of social media because of its significance to data analytics. This has interestingly attracted researchers in industry and academia to leverage the Big data techniques to benefit data analysis tools. The decision to incorporate Big data technology (i.e. Hadoop/MapReduce) in trustworthiness social data analysis is because social media content is huge and needs an efficient and scalable technology to manage it so that the data volume dimension is properly addressed. Moreover, recent literature has considered Social Networks as a form of Big data regarding volume (billions of social links), velocity (massive amount of generated content), and variety (videos, posts, mobile tweets, etc) [68], [69], [70], [71] and [72] list the main directions for BI over Big data. Shroff et al. [71] identify the importance of Big data by highlighting the effect of its characteristics on the BI domain. Big data in this context come from the data derived from the Social Networks which is unstructured and where BI tools do not have the capabilities to handle. In this context, authors of [71] show three use-cases where social-contents affect business intelligence applications dramatically: Supply-Chain Disruptions, Voice of Customer and Competitive Intelligence. However, Trust and its impact in the socio-business analysis were not addressed by the paper. Authors of [72] sum up the main areas related to Big data and BI analysis; the paper lists the evolution of BI and Analytics, their application and the research opportunities. It implements various research frameworks including Big data analytics, text analytics, web analytics, network analytics and mobile analytics. Cuzzocrea et al. [70] initiate the future research trends in the area of DW/OLAP and big data. The papers listed the main directions for the area of building and designing DW-OLAP over Big data: 1. A methodology for designing OLAP that is capable of processing Big data; 2. Efficient and complex paradigms to build OLAP's cubes over Big data; and 3. Building semantically Big data cubes. Lim et al. [69] list the main direction for the future research in the BI 2.0 in term of Big data, Text Analysis and Network analysis. Saha and Srivastava [73] present a summary to address the data veracity issue related to Big data. Poor data quality has a major negative impact on the data analysis process, and the output will lack credibility and trustworthiness. The paper addresses the data quality issues and provides tools and solutions for data in various forms (relational, structured and semi-structured); however, the unstructured data types were not addressed. Moreover, hybrid approaches could be used that utilise Ontology for data quality and trust inference purposes. The sentiment analysis of Big data is now a hot topic. Khuc et al. [74] propose a methodology for sentiment analysis that incorporates Big data technology (MapReduce/Hadoop) to process huge volumes of tweets. Although their solution addresses certain challenges, the issue of domain-based trust was omitted in their proposed approach; incorporating the notion of social trust will effectively increase the credibility of the sentiment extracted from tweets. To sum up, the research direction in the era of Big data analytics include and are not limited to: incorporating Big data technology (i.e. Map/Reduce, Hadoop) to benefit data analytics tools, developing methods to handle data in motion(real-time) for social data analytics and BI analysis, methodologies for designing OLAP tools innovatively to be capable to process SBD, and measures its credibility, and building semantically Big data cubes. In addition, starting from the characteristics of Big data and sorting out issues related to these dimensions will be the most efficient way to address Big data as well as benefit the efforts of social data analytics and the expected outcomes of Big data Analysis.

### 3.3.2 Domain-based/ Topic-Specific Trustworthiness Approaches

Adding a user-domain dimension when calculating trust in social media is an important factor. This helps to enhance understanding of users' interest. In this context, the notion of domain-based trust for the data extracted from the unstructured content (such as social media data) is significant. This is through calculating credibility values which correspond to a particular user in a particular domain. The literature of trust in social media shows a lack of approaches for measuring domain-based trust. Several reviews have carried out to highlight the importance of conducting a fine-grained trustworthiness analysis in the context of SBD [15, 75-77]. In particular, measuring the user's trustworthiness in each domain of knowledge is vital to better understanding users' behaviours in the online social services. The ontology represents the core of the domain where the knowledge is shared amongst different entities within the system that may include people or software agents [78]. In this context, a thread of efforts have been directed to a fine-grain trustworthiness analysis [7, 51, 79-83] [84]. An approach for microblogging ranking is proposed by Kuang et al. [85]. The authors incorporate three dimensions in their ranking technique (i.e. tweet popularity, the closeness between the tweet and the owner user, and the topics of interest. Zhao et al. [86] propose a scalable trustworthiness inference module for Twitterers and their tweets that take into account the heterogeneous contextual properties. Another host of scholars have addressed the issue of influential users in online social services [56, 80, 87, 88] [89]. Authors of [39] present a method to discover experts in topic-specific authority networks. They applied a modified version of the HITS Algorithm for more topic-specific network analysis. However, attributes such as (followers/following/friends counts, likes/favourites counts, etc.) are not addressed to infer user reliability. Herzig et al. [90] present an Author-Reader Influence (ARI) model that estimates a user content's attraction (i.e. content's uniqueness and relevance). In [91] the paper addresses the problem of selecting top-k expert users in the social group based on their knowledge about a given topic. In [82], the authors build a model to discover popular topics by analysing users' relationships and their interests. Jiyeon and Sung-Hyon [92] analyse the flow of information amongst users of social networks to discover "dedicators" who influence others by their ideas and specific topics. One of the top cited works in topic-based user ranking is Twitterrank [37]. Authors of Twitterrank incorporate topic-sensitive PageRank to infer topic-specific influential users of Twitter.

### 3.3.3 Assessment of Approaches Incorporating Trust in SBD

The subsections above present a review of several key techniques and approaches implemented to define and formulate the trustworthiness of users and/or their content in OSNs. These approaches follow two foremost directions. Firstly, the methods conducted to evaluate the social trustworthiness in general. These approaches address the problem of social trust and credibility in the OSNs with a lack of consideration to the users' domain of interests. Secondly, the methods contemplate the textual content of users to infer their topic(s)/domain(s) of interest first, then formulate their credibility in each topic/domain [7, 49, 83, 89]. These methods represent an advanced version of the generic trust evaluation systems.

There are two key tactics incorporated to construct the trustworthiness formula for both approaches, namely the feature-based techniques [16, 50, 93-95], and/or graph-based techniques [39, 41, 51, 80, 96, 97]. The feature-based techniques measure the trustworthiness of users and their content through incorporate the list of key attributes which primarily encompass associated metadata of the users and their content such as #followers, #friends, #likes/favourites #retweet/share, etc. The graph-based approaches evaluate the trustworthiness of users and their content in OSNs through scrutinising their social connections, where the social trustworthiness values are propagated through the whole network of users. These techniques adopt graph propagation solutions such as PageRank, HITS Algorithm, etc.

Despite the considerable efforts conducted to untangle the social trust problem; However, there are still vital dimensions need to be addressed to consolidate the proposed approaches. This includes tracking and monitoring users trustworthiness overtime; improve the existing semantic analysis techniques toward further enhancement to the contextual understanding to the users textual content; further incorporation to the sentiment analysis methods to listen to the voice of the user's followers and their opinions toward him/her. Last but not least, the implementation of the proposed approaches should address the key feature of Big data and provides technical solutions to the massive data generated steadily and incessantly from the online social platforms.

## 3.4 Case Study on Social Credibility Analysis

In this section we present a solution to tackle the credibility of users in online social services and thereby inferring domain-speccific social influcners. **Figure 1** illustrates the proposed framework and the embodied modules. As depicted in **Figure 1**, the system architecture comprises of four main phases which are detailed in the next sections, namely; (1) Tweets Acquisition and pre-processing; (2) Feature extraction and selection; and (3) Machine learning model implementation.

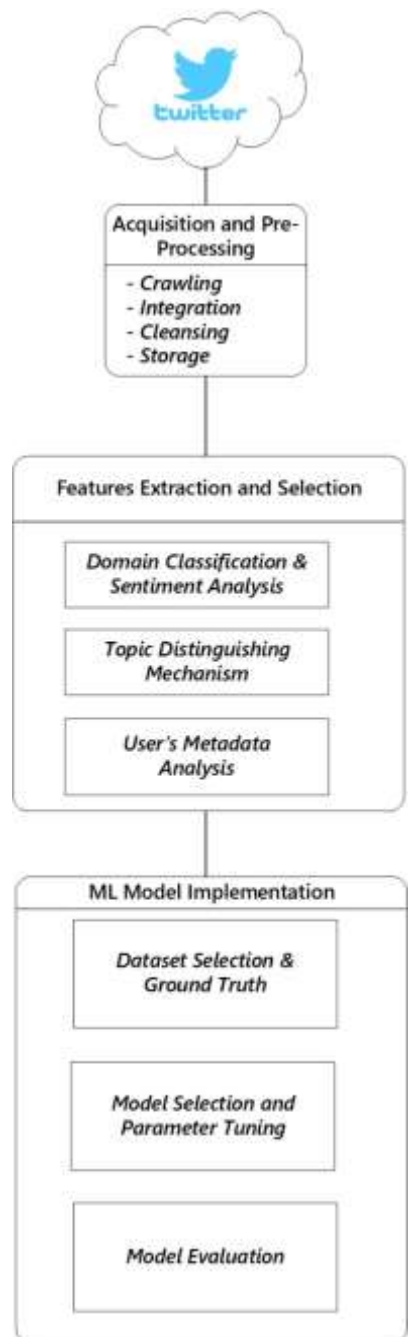

**Figure 1: System Architecture**

### 3.4.1 Tweets Acquisition and pre-processing

### 3.4.1.1 Dataset selection

This study focuses on social data generated from Twitter™. Twitter analytics is an evolving research field falling under the category of data mining, social big data and machine learning. Twitter has been chosen in this chapter due to the following reasons: (i) Twitter platform has been studied broadly in the research communities [98], leveraging the vast volume of content(6,000 tweets/seconds)[99]; (ii) It facilitates retrieving public tweets through providing APIs; (iii) the Twitter messages' "max 140 characters" feature enables data analysis and prototype implementation for a proof of concept purpose.

Twitter data access mechanisms have been harnessed in this study for data collection purposes. Users' information and their tweets and all related metadata were crawled using *TwitterAPI* [100]. A PHP script was implemented to crawl users' content and their metadata using the *User_timeline* API method. This API allows access and retrieve the collection of tweets posted by a certain user_id associated with each API request. This approach is used rather than a keyword search API due to the reasons as follows. Keyword-based search API has certain limitations listed in [98], i.e. Twitter index provides only tweets posted within 6-9 days thus it is hard to acquire historical Twitter dataset before this time span. Further, Search API retrieves results based on the relevance to the query caused by uncompleted results. This implies missing tweets and users in the search results. Using user's timeline approach, on the other hand, retrieves up to 3,200 of the recent users' tweets. Last but not least the purpose of this study is to measure the users' trustworthiness hence user-driven tweets collection is the suitable approach. Further, *acTwitterConversation*[2] API was used to retrieve all public conversations related to the tweets being fetched using Twitter API.

Data acquisition is carried out using a PHP script triggered by running a cron job which selects a new user_id and starts collecting historical user information, tweets, replies and the related metadata. The list of Twitterers' user_ids used in the data acquisition phase is extracted from a Twitter graph dataset crawled by Akcora et al. [101]. This graph is chosen since it includes the list of users who had less than 5,000 friends in 2013. This threshold was established by Akcora et al. to discover bots, spammers and robot accounts. The collection of tweets was carried out using supercomputing facilities provided and supported by Pawsey Supercomputing Centre (https://pawsey.org.au/).

### 3.4.1.2 Tweets Pre-processing

To ensure the veracity of BD, the accuracy, correctness and credibility of data should be ascertained. Although the data's origin and storage are critical to ensuring the veracity of BD, the trustworthiness of the source does not guarantee data correctness and consistency. Data cleansing and integration should also be used to guarantee the veracity of data. Further improvements to the quality of the collected data will be discussed later in the analysis phase. To address the data veracity regarding data correctness, the raw extracted tweets were subjected to a pre-processing phase. This phase includes the following steps:

---

[2] https://github.com/farmisen/acTwitterConversation

### 4.1.2.1 Data integration and temporary storage

The raw extracted tweets from TwitterAPIs are in JSON format. *AcTwitterConversation* fetches conversation tweets as an array. Data integration will provide a better-structured data format for the next analysis phase. This has been achieved by reformatting and unifying the raw tweets (JSON) and replies (ARRAY) to fit with the relational database model, the design of which is based on the metadata of the tweet, reply and the user. Then, the reformatted data is stored in a temporary location (i.e. MySql database).

### 4.1.2.2 Data Cleansing

Data at this stage may include many errors, meaningless data, irrelevant data, redundant data, etc. Thus, data is cleansed to remove noisy data and ensure data consistency. The following sequential steps were taken for the data cleansing process (i) all redundant content are eliminated (i.e. the same dataset crawled more than once) such as tweets, replies or users' data and their metadata; (ii) users who posted fewer than 50 tweets were excluded. This particular threshold was established because the aim of this research is to discover domain-based social spammers; thus, it is assumed that those users post a relatively large number of tweets in dissimilar domains of interest; (iii) media *URLs* are eliminated such as photos uploaded to Twitter, or media uploaded to one of the popular media sharing websites listed in [102] such as Instagram, Flickr, YouTube, and Pinterest. This is because these have no actual text that can be extracted for further analysis. Moreover, *URLs* directing to Facebook websites are eliminated due to the restrictions that apply to the public access of their content.

### 4.1.2.3 Data Storage

Data storage is the third phase of the BD lifecycle. *Volume* is an essential dimension to be considered when describing BD. It refers to the vast increase in the data growth where proper tools and techniques are required to manage such huge blocks of data. The data stored in this chain provides distributed and parallel data processing infrastructure based on the *Hadoop*[3] platform for Big Data. Hadoop is a distributed computing platform for data processing. It is an open source project developed by Apache[TM] to provide scalable, reliable and fault tolerant framework for Big Data. The BD infrastructure at the School of Information Systems, Curtin University, is utilised for data storage. This is a 6-node BD cluster, each with 64 GB RAM, 2 TB Storage, and 8 Core Processors. The temporal-temporary data dumps its contents to this distributed environment after the data integration process. Several

---

[3] https://hadoop.apache.org/

*Hive⁴* tables are designed and implemented in this distributed environment. The data dumps are stored in Hive tables to facilitate data access and manipulation through the SQL-like format.

### 3.4.2 Domain Classification & Sentiment Analysis

*Variability* [17] is an important BD dimension. Variability refers to variance in meaning. Incorporating semantic analysis will reduce the ambiguity of the BD. This mitigates its variability [103, 104], distinguishes users' domains of interest, and infers their genuine sentiments. The cleansed and pre-processed datasets pass through the domain classification and inference module. In this module, the textual content extracted from the datasets are analysed using two main approaches;

**Textual classification using 20 Newsgroups data set:** The 20 Newsgroups data set is a pool of around 20,000 collected documents which are classified into 20 high level categories (newsgroups) [105]. Each document in the corpus is labelled with one news category. This dataset is a popular dataset that is commonly incorporated to conduct document clustering tasks.

**Textual classification using IBM Watson – Natural Language Understanding:** is a cloud-based service which is used to extract metadata from textual content such as entities, taxonomies/categories, sentiments, and other NLP components. IBM Watson analyses the given text or URL and categorises the content of the text or webpage according to various domains/categories/taxonomies with the corresponding *scores* values. *Scores* are calculated using IBM Watson, range from "0" to "1", and convey the correct degree of an assigned category to the processed text or webpage. IBM Watson is used further to identify the overall positive or negative sentiment of the provided text. Table 1 shows an example of incorporating this API to extract categories and the sentiment of the provided tweet. As illustrated in Table 1, the content of the tweet is analysed by IBM Watson using two main modules: Categories Inference and Sentiment Analysis. The scores are provided for each module to represent the relevancy of the retrieved category and sentiment to the provided tweets. The categories inference module is used in this research in the domain discovery process, while sentiment analysis is used to discover the sentiments of tweets' replies as discussed later.

**Table 1: An example of incorporating IBM Watson for taxonomies inference and sentiment analysis**

| Tweet | | "Achieved a new personal record with @RunKeeper: Longest duration in a week #FitnessAlerts" | |
|---|---|---|---|
| Categories Inference | | Sentiment Analysis | |
| Category | Score | Sentiment | Score |
| /sports/football | 0.694 | positive | 0.51 |
| /art and entertainment/shows and events | 0.621 | | |

---

⁴ https://hive.apache.org/

| Tweet | "Achieved a new personal record with @RunKeeper: Longest duration in a week #FitnessAlerts" | | |
|---|---|---|---|
| /health and fitness/weight loss | 0.595 | | |

A tweet's content has one or two main components: *text* and *URL*. Due to the limitation of a tweet's length, a normal or legitimate Twitterer attaches with his/her tweet a URL to a particular webpage, photo, or video to help her followers obtain further information on the tweet's topic. Twitter scans URLs against a list of potentially harmful websites, then URLs are shortened using *t.co service* to maximise the use of the tweet's length. Anomalous users such as spammers abuse this feature by hijacking trends, using unsolicited mentions, etc., to attach misleading URLs to their tweets. Thus, it is important to study the tweet's domain and the comprised URL's domain to obtain a better understanding of the user's domain(s) of knowledge, which is then used to measure users' credibility.

IBM Watson is utilised further in this study to derive the sentiment of a given reply whether it is positive, negative or natural with the corresponding sentiment score. Consequently, all of a tweet's set of replies are crawled and the sentiments of these replies are incorporated in the analysis to enhance the credibility process as discussed later.

### 3.4.3 Features Extraction and Selection

The key challenge for BD analysis is the mining of enormous amounts of data in the quest for added value. *Value* of BD [106] measures the quality and significance of data with new insights. Acquiring substantial and valuable information from data in big data scale is a vital task. Researchers are trying to capture the *value* of BD in dissimilar contexts. In online social services, understanding the users' behaviour is significant due to the dramatic increase in the usage of online social platforms. This indicates the importance of measuring the users' trustworthiness, thereby discovering users' influence in a particular domain as well as discovering users with anomalous behaviour. In this study, a domain-based analysis of users' credibility is discussed to provide a comprehensive and scalable framework. This is achieved by analysing the collection of a user's tweets to measure the initial user's credibility value based on the user's historical data. This is done through the time-aware, domain-based user credibility ranking approach. The features are collected by examining the following three key techniques:

### 3.4.3.1 Topic Distinguishing Mechanism

The analysis of a user's content to discover his/her main domains of interest is an essential start to the process of measuring the user's credibility. In online social services, a user $u$ achieves a higher weight value in a certain domain(s) of knowledge if $u$ shows a sturdy interest in these domains through the posted tweets and attached URLs. This weight should be higher than those of other users who posted content in a broad range of domains. This is because no user could be conversant with all domains of knowledge [107]. Therefore, the theoretical notion of Term Frequency-Inverse Document Frequency (TF-IDF) has been used to distinguish domain-based users of online social services from others [108].

**Term Frequency-Inverse Document Frequency (TF-IDF):** TF-IDF is considered as a core component embodied onto the vector space model (VSM)[109] which is one of the classical approaches to Information Retrieval statistical models. "The intuition was that a query term which occurs in many documents is not a good discriminator"[110]. This implies that a term which occurs in many documents decreases its weight in general as this term does not show the particular document of interest to the user [111]. TF-IDF measures the importance or significance of a term to a certain document exists in a corpus of documents. It comprises standard notions which formulate its structure; Term Frequency (TF): is used to compute the number of times that a term appears in a document. TF expresses the importance of the term in the document; Document Frequency (DF): is a statistical measure to evaluate the importance of a term to a document in a corpus of texts [112]. Inverse Document Frequency (IDF) is a discriminating measure for a term in the text collection. It was proposed in 1972 [113] as a cornerstone of term weighting, and a core component of TF_IDF. It is used as a discriminating measure to infer the term's importance in a certain document(s) [114]. TF_IDF combines the definitions of TF (the importance of each index term in the document) and IDF(the importance of the index term in the text collection), to produce a composite weight for each term in each document. It assigns to a word $t$ a weight in document $d$ that is:

- Highest when $t$ occurs many times within a few number of documents.
- Lower when the term $t$ occurs fewer times in a document $d$, or occurs in many documents.
- Lowest when the term $t$ exists all documents.

In the context of this research, this heuristic aspect can be incorporated into a model to evaluate the trustworthiness of users. Consequently, it is argued that a user who posts in all domains has a low trustworthiness value in general. This argument is justified based on the following facts: (i) No one person is an expert in all domains [107]; (ii) A user who posts in all domains does not declare to other users which domain(s) s/he is interested in. A user shows to other users which domain s/he is interested in by posting a wide range of contents in that particular domain; (iii) There is the possibility that this user is a spammer due to the behaviour of spammers posting tweets about multiple topics [19]. This could end up by tweets being posted in all domains which do not reflect a legitimate user's behaviour. In other words, a user $u$ whose posts in general discuss a particular domain(s), $u$ gets a higher distinguishing value in this domain(s) and overcomes other users who usually post in a broad range of domains. This involves studying the content of users' tweets and their embedded URLs to obtain a thorough understanding of their domain(s) of knowledge as it will be elaborated through this section.

### 3.4.3.2 Users' Metadata Analysis

It is important to have an understanding of the interactions-based attributes of OSN users, as this is a significant factor when discovering socially reliable, domain-based users as well as inferring users with anomalous behavior. This involves studying all the related metadata extracted from users' content(#likes, #replies, #retweets, #friends, #followers, etc.). This also includes the followers' interest in the users' content, their positive or negative opinions, etc.

### 3.4.3.3 Features Extraction

A metric incorporating several key attributes are used to build the feature-based ranking model. **Table 2** provides a summery to the incorporated features extracted from both users' data and their metadata. A detailed discussion and explanation to most of these features is provided in our previous works [29, 35].

**Table 2: Features description**

| Feature | Description | Equation |
|---------|-------------|----------|
| Tweet Similarity Penalty (**Twt_Sim**) | Represents the count of unique keywords (**#distinct words**) in the overall user's tweets to the total number of keywords in the user's tweet (**#words**). | $Twt\_Sim_u = \dfrac{\#DistinctWords_u}{\#Words_u}$ |
| URL Similarity Penalty (**URL_Sim**) | Represents the percentage of non-redundant URLs (**#DistinctURLs**) with non-redundant hosts of URLs (**#DistinctURLsHosts**) to the total number of URLs (**#URLs**) posted by user **u**. | $URL_{Sim_u} = 0.5 \times \left( \dfrac{\#DistinctURLs_u + \#DistinctURLsHosts_u}{\#URLs_u} \right)$ |
| Domain based content user score (**Sum_cnt_scr**) | Is computed by adding all scores retrieved from IBM Watson of tweets' texts posted by user **u** in domain **d**. | - |
| Domain based user URL scores (**Sum_url_scr**) | Is calculated by accumulating scores for all websites' content of the URLs embedded in user **u**'s tweets in domain **d**. | - |
| Domain based user scores (**Sum_all_scr**) | Refinement summing of the corresponding scores achieved by IBM Watson for all tweets' texts ($Sc_{u,d}^{Twt}$), and the refinement summing of scores retrieved from URLs' webpage content ($Sc_{u,d}^{URL}$) posted by a user **u** where a domain **d** was inferred | $Sc_{u,d} = (Twt\_Sim_u \times Sum\_cnt\_scr_{u,d}^{Twt} + URL\_Sim_u \times Sum\_url\_scr_{u,d}^{URL})$ |
| Domain frequency (**DF**) | Count of domains the user **u** has tweeted about. | - |
| Inverse domain frequency (**IDF**) | Distinguishes users among the list of their domains of interest. | $IDF_u = log(\dfrac{n}{DF_u})$ |
| Weight (**W**) | Users weights in each domain. | $W_{u,d} = Sc_{u,d} \times IDF_u$ |
| Domain-based user's retweets (**R**) | Represents the frequency of retweets for user' content in each domain **d** | - |
| Domain-based user's likes (**L**) | Represents the percentage of likes/Favourites count for the users' content in each domain **d** | - |
| Domain-based user's replies (**P**) | Embodies the count of replies to the users' content in each domain **d** | - |

| Feature | Description | Equation |
|---|---|---|
| Domain-based user positive sentiment replies ($SP$) | Refers to the sum of the positive scores of all replies to a user $u$ in domain $d$. Positive scores are achieved from IBM Watson with values greater than "0" and less than or equal to "1". The higher the positive score, the greater is positive attitude the repliers have to the users' content. | - |
| Domain-based user negative sentiment replies ($SN$) | Represents the sum of the negative scores of all replies to a user u in domain d. Negative scores are those values greater than or equal to "-1" and less than "0". The lower the negative score, the greater is the negative attitude the repliers have to the users' content. | - |
| domain-based user sentiments replies ($S$) | Embodies the difference between the positive and negative sentiments of all replies to user $u$ in the domain $d$. | $S_{u,d} = SP_{u,d} - \lvert SN_{u,d} \rvert$ |
| Users' followers ($FOL$) | Total count of users' followers. | - |
| User's friends $FRD$ | Total count of user's friends (followees) | - |
| Followers-friends ratio. $FF\_R$ | User followers-friends ratio. | $FF\_R_u = \begin{cases} \frac{FOL_u - FRD_u}{Age_u}, & if\ FOL_u - FRD_u \neq 0 \\ \frac{1}{Age_u}, & if\ FOL_u - FRD_u = 0 \end{cases}$ , <br> where $Age$ is the age of user profile in years |

### 3.4.4 Experimental Results

### 3.4.4.1 Dataset selection and Ground Truth

To evaluate the credibility of users incorporating the temporal factor, the cleansed dataset is divided into six chunks, where each chunk is comprised of the data and metadata of each particular month. These chunks embody the chronologically sequential snapshots indicating the recent user's activity amongst the crawled dataset. **Figure 2** shows the total count of *users*, *tweets* and their *replies* for the determined time. The number of users shown in **Figure 2** (i.e. 6,066) represents the total distinct number of users who posted tweets in one or more of the determined period. The remaining users posted their tweets before that, although they have been inactive in twitter recently. This signifies the importance of studying users' content temporally.

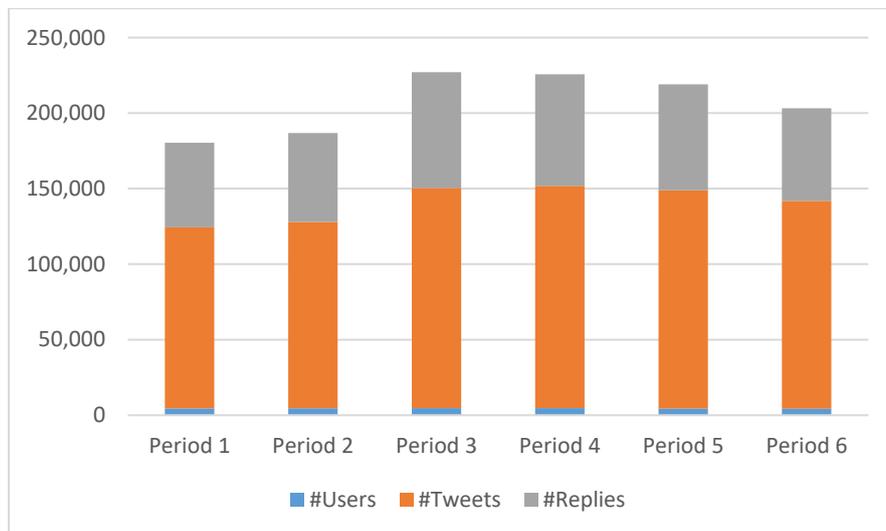

**Figure 2: Total monthly count of users, tweets and replies**

### 3.4.4.2 System Evaluation

Due to the size of chapter, we have selected "Technology and Computing" domain and conducted labelling on more than 4000 extracted users to classify them into two categories, namely; *Influence* and *non-Influence* in "Technology and Computing" domain. This experiment has been conducted using RapidMiner™ software. **Figure 3** shows the number of influencers in IT in a comparison with the number of non-influencers in IT domain.

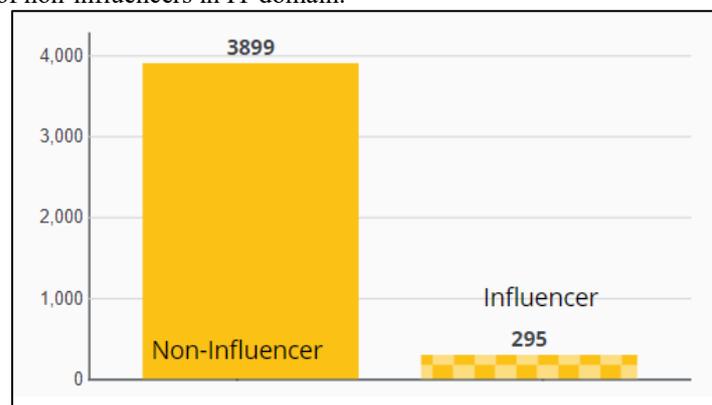

**Figure 3: Number of influencers and Non-influencers in IT domain**

As indicated in **Figure 3**, the numbers of influencer users are less than non-influencer users. This is due to the fact that users might be legitimate and trustworthy in a certain domain of knowledge this, but this does not indicate their influence in this designated domain. Users should how high level of knowledge acquisition and expertise to be indicated as knowledge-based influencers.

**Figure 4** shows the correlation of number of users between the calculated trustworthiness values in the "Technology and Computing" domain and each selected feature. It is evident that number of users who obtained the high credibility value in IT domain have gained high value in each designated feature depicted in **Figure 4**. This indeed supports the facts illustrated in **Figure 3** where numbers of influencers are relatively less than numbers of non-influencer users.

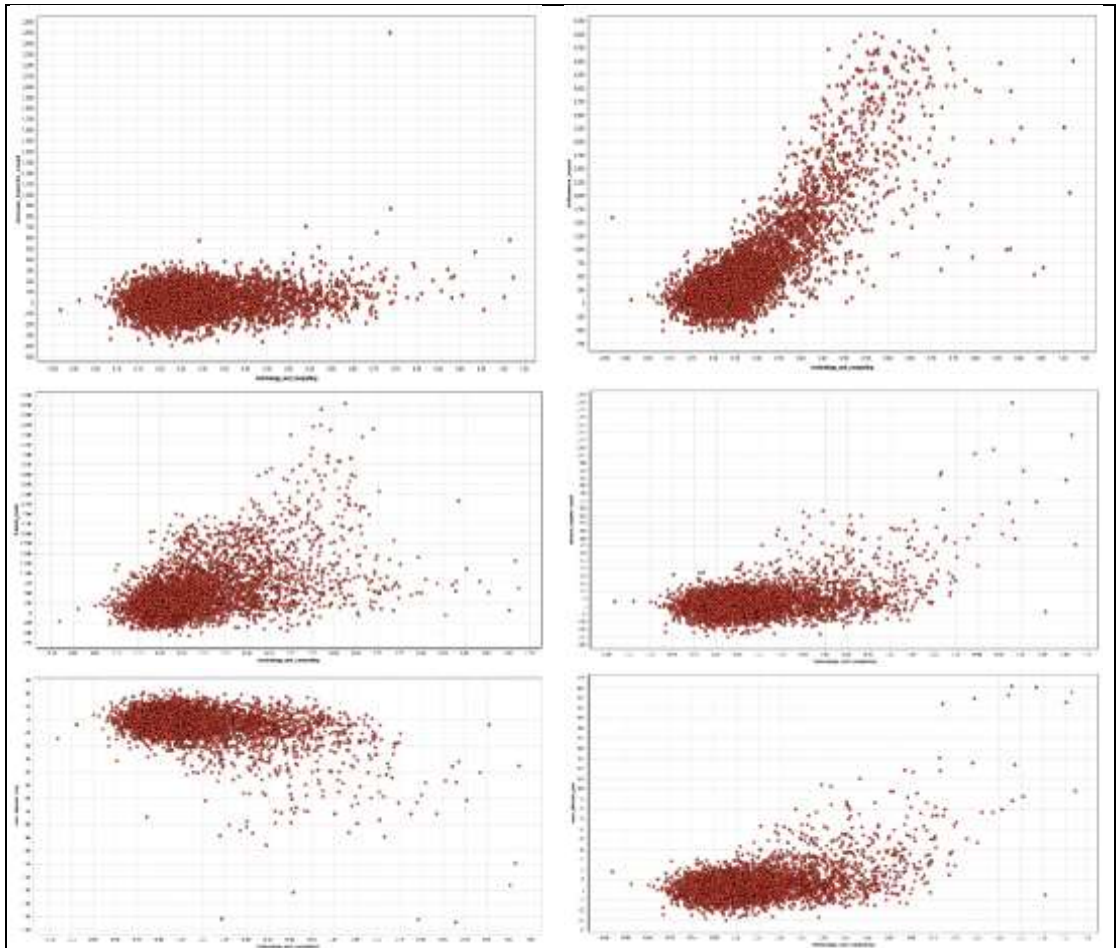

**Figure 4: correlation between the trustworthiness value in IT domain and each feature**

**Figure 5** and Figure 6 show the accuracy of the conducted prediction task and the prediction errors in each module respectively. It is evident that deep learning prediction module has achieved the highest prediction estimates along with the logistic regression. This indicates the adequacy of logistic regression in conducting classification and prediction tasks that might not require high level of computational prediction modules such as deep learning techniques.

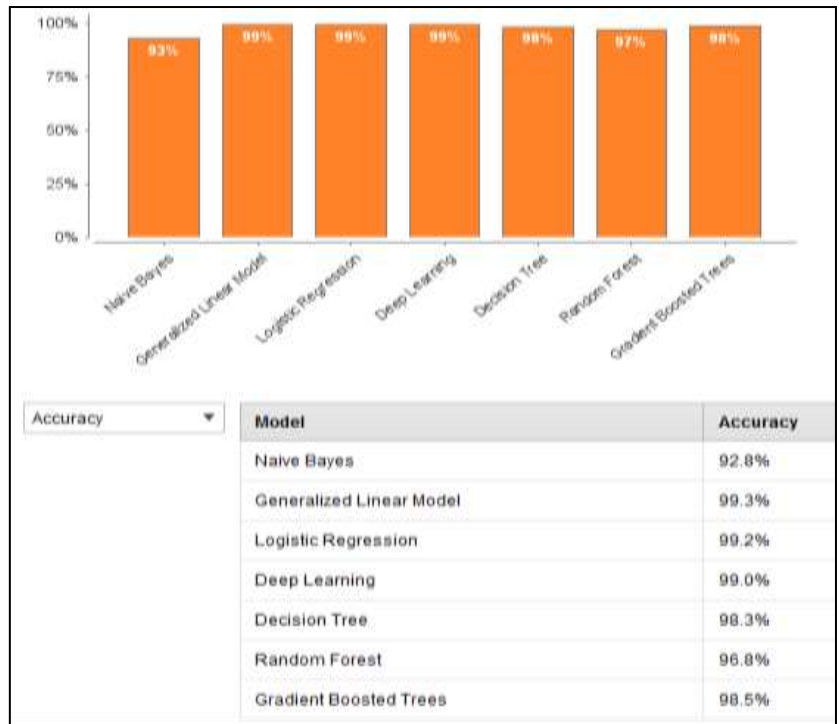

| Accuracy ▼ | Model | Accuracy |
|---|---|---|
| | Naïve Bayes | 92.8% |
| | Generalized Linear Model | 99.3% |
| | Logistic Regression | 99.2% |
| | Deep Learning | 99.0% |
| | Decision Tree | 98.3% |
| | Random Forest | 96.8% |
| | Gradient Boosted Trees | 98.5% |

**Figure 5: Accuracy of the proposed framework**

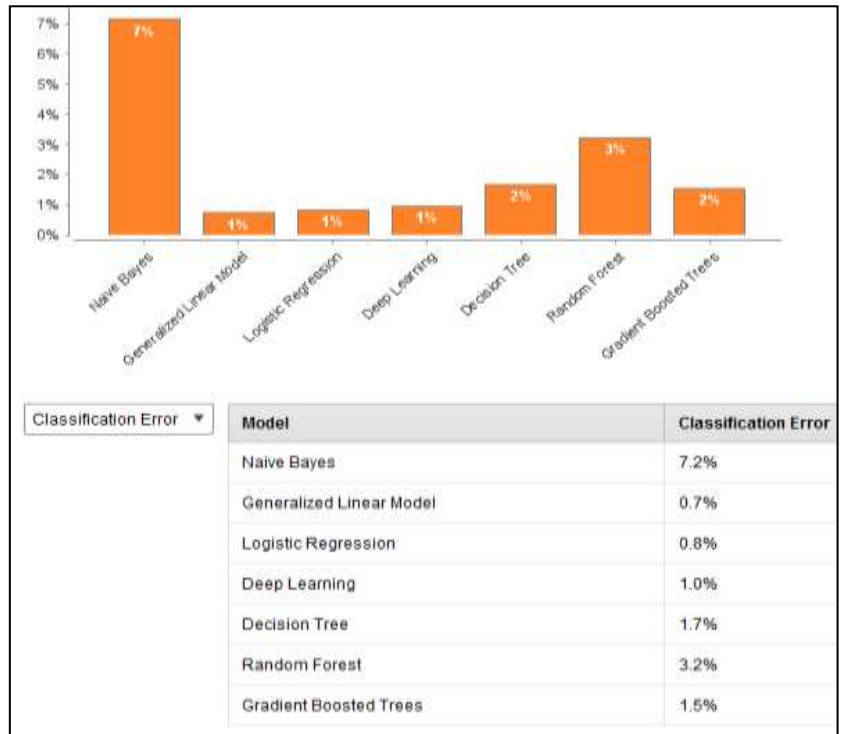

**Figure 6: Classification Error of each prediction module**

| Classification Error ▼ | Model | Classification Error |
|---|---|---|
| | Naïve Bayes | 7.2% |
| | Generalized Linear Model | 0.7% |
| | Logistic Regression | 0.8% |
| | Deep Learning | 1.0% |
| | Decision Tree | 1.7% |
| | Random Forest | 3.2% |
| | Gradient Boosted Trees | 1.5% |

**Figure 7** shows the ROC curves for all models, together on one chart. The closer a curve is to the top left corner, the better the model is. As depicted in Figure 7 logistic regression and deep learning have shown adequacy in the conducted prediction task.

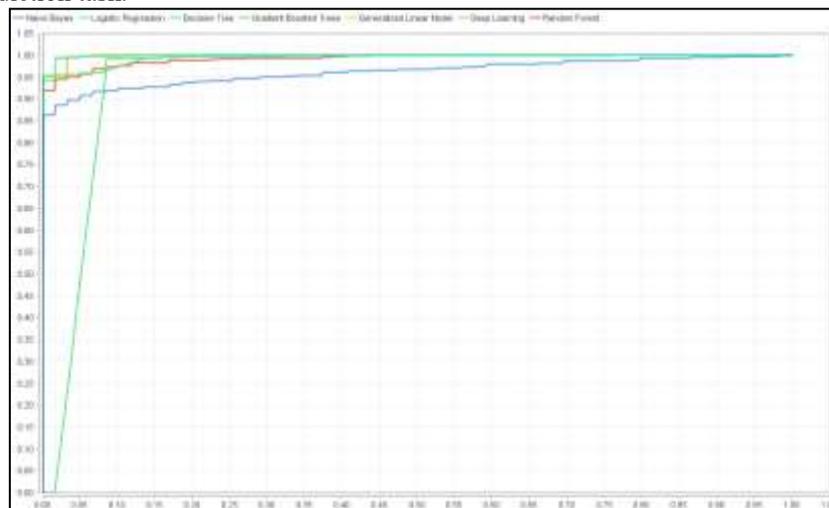

**Figure 7: ROC Curve of the incorporated prediction modules**

## 3.5 Conclusion

This chapter presents an overview depiction to the credibility problem in SBD. First, a definition to the credibility is given followed by providing various approaches used to tackle the generic and domain-specific credibility. Finally, an approach to estimate and predict the domain based credibility in online social services is proposed as a case study. The experimental task conducted to evaluate this approach validates the applicability and effectiveness of indicating influencers and non-influencers users in the designated domain. The developed approaches in this chapter have produced optimistic results. However, there are certain limitations that need to be addressed and possible enhancements to be elucidated and marked as future work; (i) IBM Watson has been used in this framework as the sole semantics provider. The resultant semantics should be enhanced further by utilising an ontology-based approach; (ii) a new graph-based model will be created to propagate the users' credibility throughout the entire network. Hence, an enhanced version of Twitterrank [37] is anticipated that takes into consideration the semantics of the textual content and the temporal factor and (iii) an anomaly detection approach will be developed that incorporates machine learning and an advanced list of features.